\documentstyle[aps,preprint]{revtex}
\begin{document}
\draft
\def\zt{Zlatko Te\v sanovi\' c}
\title{\bf Density of States of a Type-II Superconductor in 
a High Magnetic Field: Impurity Effects 
 }
\newcommand{\q}{\vec{q}} 
\newcommand{\r}{\vec{r}}
\newcommand{\k}{\vec{k}}
\newcommand{\be}{\begin{equation}}
\newcommand{\ee}{\end{equation}}
\author{
 Sa\v {s}a Dukan$^{a}$ 
 and Zlatko Te\v {s}anovi\' {c}$^b$}
\address{
\sl a) Department of Physics and Astronomy, Goucher College, Baltimore, MD 21204\\
\sl b) Department of Physics and Astronomy, 
Johns Hopkins University,
Baltimore, MD 21218} 
\maketitle

\begin{abstract}
~~We have calculated the density of states ${\cal N}(\omega )$ of a dirty
but homogeneous superconductor in a high magnetic field. We assume a 
dilute concentration of scalar impurities and find how ${\cal N}(\omega )$
behaves as one crosses from the weak scattering to the strong scattering limit.  
At low energies, ${\cal N}(\omega )\sim \omega ^2$ for small values
of the impurity concentration and scattering strength. When the disorder
becomes stronger than some critical value, a finite density of states is 
created at the Fermi surface. 
These results are a consequence of the gapless nature of the quasiparticle excitation
spectrum in a high magnetic field.

\end{abstract}
\narrowtext

\section{MOTIVATION}

When an electronic system is placed in an external magnetic field, motion
of electrons perpendicular to the field direction is confined to 
cyclotron orbits. The electronic energies are quantized in the form of a discrete
set of  
Landau levels (LL's) separated by $\hbar 
\omega _c$, where $\omega _c=eH/mc$ is the cyclotron frequency.
In high magnetic fields and at low temperatures, LL quantization
leads to unusual behavior in many-body systems. Perhaps the most vivid
example of such behavior is the fractional quantum
Hall effect in a two-dimensional (2D) Fermi system subjected to a strong perpendicular
magnetic field. There has also been much interest lately in the influence of 
LL quantization on properties of superconductors in a magnetic field. It
was shown by Te\v{s}anovi\'{c} et.al. \cite{zbt}, that standard Abrikosov-Gor'kov
theory of type-II superconductors which neglects LL quantization, necessarily breaks 
down at high fields and low temperatures. The inclusion 
of LL's in the BCS description of the superconducting instability leads to 
reentrant behavior at high fields (much larger than the semi-classical upper
critical field $H_{c2}(T)$) where the superconductivity is
enhanced by a magnetic field \cite{zbt}.  
As a consequence of the underlying LL
structure, $H_{c2}(T)$, or rather $T_c(H)$, develops 
oscillations near $H_{c2}(0)$. Similar types of quantum oscillations, with
the same origin, have been predicted in various other measurable quantities
which are particularly pronounced in 2D systems \cite{maniv}.
An important theoretical question in this context is the nature of
the quasiparticle excitation spectrum at high magnetic fields. It is well 
known that the existence (or, typically, absence) of low-energy excitations in the 
superconducting state shapes the behavior of all superconducting thermodynamic and transport 
properties. The problem of quasiparticle excitations has been 
studied extensively in the low field limit ($H\geq H_{c1}(T)$) in the context
of an isolated vortex line. Scanning tunneling experiments 
\cite{hess} have revealed the structure of bound states in the
cores of isolated vortices at low fields, thereby confirming an early
prediction of localized midgap states by Caroli, de Gennes and Matricon \cite{caroli}.
Only recently, was the problem of quasiparticle excitations in 
high magnetic field in the presence of a vortex lattice addressed by a
number of groups \cite{sasa,stephen,norman,oxford}: At high 
magnetic fields in extreme type-II superconductors 
vortices are closely packed ($H\leq H_{c2}(T)$) so that the isolated vortex picture 
from the low-field limit necessarily breaks down. In clean samples at low temperatures 
quasiparticles can propagate coherently over many unit cells of the 
vortex lattice. The solution of the BCS problem \cite{sasa}, where
the coherent nature of quasiparticle excitations is fully accounted for,
points to the qualitatively new nature of the energy spectrum at high 
fields: at fields near $H_{c2}$ this spectrum is {\it gapless} at a discrete set of points
on the Fermi surface. This gapless behavior in 3D systems persists to a surprisingly low
magnetic field $\sim 0.5H_{c2}$ \cite{thesis}, 
below which the gaps start opening up, and the system eventually reaches the regime of localized
states in the cores of isolated vortices \cite{norman1}. The gapless character of the excitation spectrum leads to algebraic behavior 
of various thermodynamic functions in the high field/low temperatures portion 
of the H-T phase diagram \cite{sasa}. In particular, it was shown by Dukan and Te\v{s}anovi\'c \cite{prl} that
the recent observations of the dHvA-oscillations in the mixed state of A-15
superconductors \cite{corcoran,spring} follow from the presence
of a small portion of the Fermi surface containing gapless quasiparticle
excitations, surrounded by regions where the BCS gap is large. This picture
has received further support in the dHvA experiments on YNi$_2$B$_2$C \cite{gall}. 

In this paper we address the influence of the non-magnetic (scalar) impurities on the 
gapless superconducting state in high fields, by
examining the behavior of the 
superconducting density of states (DOS) in the presence of disorder. We are interested in a {\it dirty} but
{\it homogeneous} superconductor for which the coherence length $\xi $ is 
much longer then the effective range $\xi _{imp}$ of the impurity potential. Under 
this condition the order parameter in the mixed state is not substantially affected
by the impurities and it is assumed to form a perfect triangular vortex lattice.  
In the opposite, inhomogeneous, limit the impurity potential can pin the vortex 
making an otherwise perfect vortex lattice disordered. The problem of 
the superconducting instability and the quasiparticle excitation spectrum in
the presence of randomly distributed vortex lines has been addressed so far only by Gedik and Te\v{s}anovi\'{c} 
in the quantum limit, where electrons occupy only the lowest LL. 
It was  demonstrated that the gaplessness of the excitation spectrum found in a 
pure superconductor is preserved even when the positions of the vortex lines 
are completely random \cite{gedik}.

 To understand the effect of disorder on the superconducting properties in 
the high-field limit near $H_{c2}(0)$ we start with the normal state in this limit. This
problem has been thoroughly investigated in the study 
of transport properties in high fields \cite{abrikosov3}, where it was found that the electronic scattering with scalar point-like impurities randomly 
distributed through the sample leads to isotropic broadening by $\delta 
E=\hbar /2\tau $ of the (LL's), where $1/2\tau$ is the scattering rate due to the 
disorder. As long as the broadening $\delta E\ll \hbar \omega _c$, the 
discreteness of the LL structure is preserved and the effect of the disorder 
on the superconducting state can be analyzed pertubatively starting from the 
results for the pure superconductor in high field. 
We present a Green's function pertubative approach to impurity
effects in this regime based on the theory of superconducting alloys of Abrikosov and Gor'kov \cite{abrikosov2} 
This is a standard field-theory technique of treating the disorder in the
superconducting state, and has been used extensively in the study 
of anisotropic superconductors such as heavy-fermion systems, p- or d-wave
unconventional superconductors \cite{hirch} or in the study of superfluidity
in $^3$He films \cite{vallis}.

This paper is organized as follows: In Sec. II we describe a simple model
of electronic scattering from point-like scalar impurities randomly
distributed in the superconducting material. In Sec. III we develop 
a self-consistent procedure for calculating the electronic self-energies
in the framework of the Self-consistent Born (SCBA) approximation. We discuss the behavior of 
the superconducting density of states within this approximation.
In Sec. IV we study the low-energy properties of the density of states 
in the self-consistent T-matrix approximation. 
In Sec. V we discuss the consequences of our results on various
experimentally measurable properties of a superconductor in high magnetic
field. 

\section{DESCRIPTION OF THE MODEL} 

We consider a 3-dimensional (3D) BCS-type, weak-coupling electronic system in a high magnetic field.
We assume that the system in question is translationally invariant in the normal state. The aspects
associated with the true band structure of a particular material are not important for the issues
that we are planning to address and are not discussed in this work. Nevertheless, all the results
presented here can be modified to include band structure effects. For a weakly to moderately interacting system,
one uses a simple short-range attractive BCS model interaction $V(\vec{r_1},\vec{r_2})=-V\delta (
\vec{r_1},\vec{r_2})$ between electrons. The dynamical origin of this interaction (i.e. whether mediated by phonons, charge density-fluctuations, etc.) is not 
important for our present purposes. We assume that the presence of impurities does not 
effect the effective electron-electron interaction so that it is still of the BCS form. 
Furthermore, the coupling constant $V$ is taken to be independent of 
magnetic field. The mean-field (MF) Hamiltonian for this model system is:
\begin{eqnarray}
H=\sum_{\alpha,\beta =1,2}\int \Psi _{\alpha}^{\dagger}
(\vec {r})[\frac{1}{2m}(-i\hbar \nabla +\frac{e}{c}\vec{A})^{2}\delta _{\alpha \beta }+U_{\alpha \beta}(\vec{r})-g\mu _B\vec{\sigma }\cdot \vec{H}(\vec{r})-\mu ]\Psi _{\beta }({\bf r})d^{3}r+
\nonumber\\
\int \Delta(\vec{r})\Psi _{\uparrow}^{\dagger}(\vec{r})
\Psi _{\downarrow}^{\dagger}(\vec{r})d^{3}r+h.c.
\label{hf}
\end{eqnarray}
where $\Psi _{\alpha}(\vec{r})$ are the electron field operators for two spin components and $\mu $ is chemical
potential. $\vec{A}$ is the vector potential due to the external field $\vec{H}$, which is taken to be uniform everywhere in the system.  
Term $-g\mu_B\vec{\sigma}\cdot\vec{H}$ describes the Zeeman splitting and $U_{\alpha \beta }(\vec{r})=\sum
_jU_{\alpha \beta}(\vec{r}-\vec{R}_{j}^{imp})$ is the random impurity contribution. $\Delta (\vec{r})=
V< \Psi _{\uparrow}^{\dagger}(\vec{r})
\Psi _{\downarrow}^{\dagger}(\vec{r})>$, with $<...>$ denoting thermodynamic average,  
is
the superconducting order parameter.

For simplicity we 
consider a system in which Zeeman splitting is negligible, i.e. $g\approx 0$.
The results are straightforwardly generalized to the $g\approx 2$ case \cite
{sasa,stephen,norman,oxford,thesis}.
 We assume that there are only non-magnetic (scalar) impurities present in the
sample so that $U_{\alpha \beta}(\vec{r})$ does not contain
any spin-exchange terms. We consider dilute impurity concentrations for which
the MF picture presented in this paper is valid. For a {\it dirty} but
{\it homogeneous} superconductor in which the coherence length $\xi $ is 
much longer than the effective distance $\xi _{imp}$ over which the impurity 
potential changes i.e., $\xi /\xi_{imp}\gg 1$, the superconducting order parameter
$\Delta (\vec{r})$ in Eq. (1) is not affected by the impurities apart from its 
overall magnitude, and forms a perfect 
triangular Abrikosov lattice \cite{abrikosov1}. A generic example of such an impurity potential is an infinitely short-range $\delta $-function potential of the
form:
\begin{equation}
U_{\alpha \beta}(\vec{r})=\sum_{i}U(\vec{r}-\vec{R_i})=\sum_{i}U_{o}\delta(\vec{r}-\vec{ R_{i}})
\label{imp}
\end{equation}
where $\vec{R_{i}}$ is the location of the $i$th impurity taken to be completely randomly distributed everywhere in the sample. The scalar 
scattering amplitude $U_o$ is assumed to be isotropic. 

It was shown in Ref. 5 that the unperturbed part of MF Hamiltonian (1) with $U_{\alpha \beta}(\vec{r})=0$ can be diagonalized in
terms of the basis function of the Magnetic Sublattice Representation (MSR) 
\cite{byckov}, 
characterized by the quasi-momentum ${\q}$ perpendicular to the
direction of the magnetic field. 
 The eigenfunctions of this representation 
in the Landau gauge $\vec{A}=H(-y,0,0)$ and belonging to the $m$th Landau 
level are:
\begin{eqnarray}
\phi _{k_{z},\q ,m}(\vec{r})=\frac{1}{\sqrt{2^{n}n!\sqrt{\pi }l}} \sqrt{
\frac{b_y}{L_xL_yL_z}}\exp {(ik_z\zeta )}\sum_{k}\exp {(i\frac{\pi b_x}{2a}k^2-ikq_y
b_y)}  
\nonumber\\ 
\exp {[i(q_x+\frac{\pi k}{a})x-1/2(y/l+q_xl+\frac{\pi k}{a}l)^2]}H_{m}(\frac
{y}{l}+(q_x+\frac{\pi k}{a})l).
\label{phi}
\end{eqnarray}
where $\zeta $ is the spatial coordinate and $k_{z}$ is the momentum along the 
field direction. $\vec{a}=(a,0)$ and $\vec{b}=(b_x,b_y)$ are the unit vectors 
of the triangular vortex lattice, $l=\sqrt{\hbar c/eH}$ is the magnetic
length and $L_xL_yL_z$ is the volume of the 
system. $H_{m}(x)$ is the Hermite polynomial of order $m$. 
Quasimomenta ${\q}$ are restricted to the first Magnetic Brillouin Zone
(MBZ) defined by vectors $\vec{Q}_1=(b_y/l^2,-b_x/l^2)$ and $\vec{Q}_2=
(0,2a/l^2)$. In the Landau gauge the Abrikosov order parameter can be written as:
\begin{equation}
\Delta ({\r})=\Delta \sum_n\exp{(i\pi \frac{b_x}{a}n^2)}\exp{(i2\pi nx/a-(y/l
+\pi nl/a)^2)}
\label{order}
\end{equation}
where $\Delta $ is the overall BCS amplitude. The above form of the order parameter
is taken to be entirely contained in the lowest LL of Cooper charge $2e$. 
This is an excellent approximation in the high-field regime \cite{zbt}.
Normal and anomalous Green's functions for the clean superconductor in this representation can be constructed as: 
\begin{eqnarray}
{\cal G}(\vec{r},\vec{r'};\omega )=\sum _{n,k_{z},\vec{q}}
\phi _{n,k_{z},\vec{q}}(\vec{r})\phi _{n,k_{z},\vec{q}}^{*}(\vec{r'})
G_n(k_{z},\vec{q};\omega )
\nonumber \\ 
{\cal F}^{\dagger }(\vec{r},\vec{r'};\omega )=\sum _{n,k_{z},\vec{q}}
\phi ^{*}_{n,-k_{z},-\vec{q}}(\vec{r})\phi ^{*}_{n,k_{z},\vec{q}}(\vec{r'})
F_n^*(k_{z},\vec{q};\omega )
\label{green} 
\end{eqnarray}
where $\omega =k_BT(2m+1)\pi $ are the electron Matsubara frequencies.
In writing (\ref{green}) we have taken into account only diagonal (in Landau level index $n$) contributions to the Green's functions.
(\ref{green}). This is an excellent approximation in high magnetic fields 
where $\Delta /\hbar \omega _{c} \ll 1$ and number of occupied Landau level 
$n_{c}$ is not too large. In this situation we can use 
the diagonal 
approximation  
\cite {zbt}, in which the BCS pairs are formed by the electrons belonging to 
the mutually degenerate Landau levels at the Fermi surface while the 
contribution from the Landau levels that  are separated by $\hbar \omega _c$ or
more is included in the renormalization of the effective coupling constant ($V
\rightarrow \tilde{V}(H,T))$ \cite{thesis}.   
In lower fields, where 
$n_c$ is a large number, the off-diagonal terms in (\ref{green}) should
be included on equal footing. It was shown in Ref. 5 by numerically solving the BCS equations 
that the off-diagonal pairing does not 
change the qualitative behavior of the superconductor in a magnetic field as long as 
the magnetic field is larger than some critical field $H_{critical}(T)$ 
(estimated to be $\approx 0.5H_{c2}$ at $T\approx 0$ in A-15 superconductors): Placed in a high magnetic field
and cooled to low temperatures the type-II superconductor has a gapless 
excitation spectrum. In the diagonal approximation the gapless branches in
the spectrum 
can be found analytically as:
\begin{eqnarray}
E_{n}(k_{z},{\q})=\pm \sqrt{\epsilon _{n}^{2}(k_z)+|\Delta _{nn}({\q})|^{2}}
\nonumber\\
\epsilon _{n}(k_z)=\frac{\hbar ^2k_{z}^{2}}{2m}+\hbar \omega _{c}(n+1/2)-\mu
\label{epsilon}
\end{eqnarray}
where  $\Delta _{nm}({\q})$ are the matrix elements of the Abrikosov order
parameter (\ref{order})  in the MSR representation and can be written as:
\begin{equation}
\Delta _{nm}(\q )=\frac{\Delta }{\sqrt{2}}\frac{
(-1)^{m}}{2^{n+m}\sqrt{n!m!}}\sum_kexp(i\pi \frac{b_x}{a}k^2+2ikq_yb_y-(q_x+\pi k/a)^2l^2)
H_{n+m}[\sqrt{2}(q_x+\pi k/a)l]
\label{delta}
\end{equation} 
Once the off-diagonal pairing is included, the excitation spectrum cannot be written in the  
simple form (\ref{epsilon}) and a closed analytic expression for the superconducting Green's function cannot be found.
Nevertheless, the qualitative behavior of the quasiparticle excitations, characterized by
the nodes in the MBZ, remains the same. The main role of the off-diagonal terms in
Eqs. (5) for magnetic field strengths such that $H>H_{critical}$ is to renormalize the slopes around the nodes.  
Once the magnetic field is lowered below $H_{critical}$, the gaps start
opening up at the Fermi surface signaling the crossover to the low-field
regime of quasiparticle states localized in the cores of widely separated vortices 
\cite{norman1}. 
In this paper we are interested in how the disorder affects the excitation spectrum in the gapless regime 
and,in particular, how the DOS behaves in this 
high-field (and low-temperature) gapless regime of a dirty superconductor. For this purpose, 
we argue that taking only the diagonal terms in Eqs. (5) will 
correctly capture the low-energy behavior of DOS for a wide 
range of magnetic fields (as long as $H>H_{critical}$) while considerably
reducing the computational difficulties introduced by the off-diagonal terms.

Proceeding in the frame described above, the ``Fourier transformed" (in the quasimomentum space) Green's functions $G_n^0(k_z,\vec{q};\omega )$ and
$F_n^{0*}(k_z,\vec{q};\omega )$ for the pure superconductor can be easily
calculated as:
\begin{eqnarray}
G^0_n(k_{z},\q ;\omega )=\frac{-i\omega 
 -\epsilon _{n}(k_{z})}{\omega ^{2}+E_{n}^{2}(\q ,k_{z})} 
\nonumber\\
F_n^{*0}(k_{z},\q ;\omega )=\frac{\Delta _{nn}^{\ast }(\q )}
{\omega ^{2}+E_{n}^{2}(\q ,k_{z})}
\label{f}
\end{eqnarray} 
with
$E_{n}(k_{z},\vec{q})$ and
$\Delta _{nm}(\vec{q})$ given by Eqs. (6) and (7), respectively.

\section{SELF-CONSISTENT BORN APPROXIMATION}

 The equations of motion describing the mixed superconducting state in
the presence of scalar impurities are:
\begin {eqnarray}
\left[ i\omega +\frac{1}{2m}[\nabla _{{\r}}-ie\vec{A}({\r})]^2+\mu -\sum_{i}U(\vec{r}-\vec{R_i})\right] {\cal G}(\vec{r},\vec{r'};\omega )+
\Delta(\vec{r}){\cal F}^{\dagger }(\vec{r},\vec{r'};\omega )=\delta (\vec{r}-\vec{r'}),
\nonumber\\
\left[-i\omega +\frac{1}{2m}[\nabla _{{\r}}+ie\vec{A}({\r})]^2+\mu -\sum_{i}U(\vec{r}-\vec{R_i})\right] {\cal F}^{\dagger }(\vec{r},\vec{r'};
\omega )-\Delta ^{\ast }(\vec{r}){\cal G}(\vec{r},\vec{r'};\omega )=0 
\label{abg}
\end{eqnarray}
where normal and anomalous Green's functions incorporate the interactions
between the electrons and impurities. We are interested in the quantities ${\cal G}$ and ${\cal F}^{\dagger }$ averaged over the positions of the 
impurities.

 In performing the average over the disorder we follow closely the diagrammatic  technique 
developed by Abrikosov and Gor'kov in Ref. \cite{abrikosov2}
for the study of the superconducting alloys in zero field. First, we notice
that for a dirty but homogeneous superconductor the quasi-momentum $\vec{q}$ is
still a good quantum number, since the Abrikosov lattice is not
affected by a short-ranged impurity potential (\ref{imp}). Therefore,
the Green's functions in (\ref{abg}) can be expanded in terms of a complete
set of eigenfunctions (\ref{phi}) as:
\begin{eqnarray}
{\cal G}(\vec{r},\vec{r'};\omega )=\sum_{k_1}\sum_{k_2}\phi _{k_1}(\vec{
r})\phi _{k_2}^{*}(\vec{r'})G(k_1,k_2;\omega )
\nonumber\\
{\cal F}^{\dagger }(\vec{r},\vec{r'};\omega )=\sum_{k_1}\sum_{k_2}\phi ^{*}
_{-k_1}(\vec{r})\phi ^{*}_{k_2}(\vec{r'})F^{*}(k_1,k_2;\omega ) 
\label{expansion}
\end{eqnarray}
where $k\equiv (\vec{q},k_z,n)$. 
 
The Hamiltonian for the interaction with impurities contains operator products
$\Psi \Psi ^{\dagger}$. Therefore, when an impurity is inserted into an
electron line, two possibilities arise for each of the propagators $\cal G$,
$\cal F$ and ${\cal F}^{\dagger}$. These possibilities can be written in 
matrix form:
\begin{equation} 
\hat{{\cal G}}(x,x')\rightarrow \hat{{\cal G}}(x,y)\hat{\sigma }_z\hat{{\cal G}}(y,x')
\label{pauli1}
\end{equation} 
with the $2\times 2$ matrix $\hat{{\cal G}}(x,x')$ defined in Nambu formalism, 
$\hat{\sigma}_z$ the Pauli matrix, and $x\equiv ({\r},\tau )$, where $\tau $
is the imaginary time. Taking into account the expansion (\ref{expansion})
this matrix equation can be rewritten in terms of its ``Fourier" components as:
\begin{eqnarray}
G(k_1,k_2;\omega )=G^{0}(k_1;\omega )\delta (k_1-k_2)
+G^{0}(k_1;\omega )\sum_{k_3,\vec{R}_i}U_{k_1k_3}(\vec{R}_i)
G(k_3,k_2;\omega )
\nonumber\\
-F^{0}(k_1;\omega )\sum_{k_3,\vec{R}_i}U_{-k_3-k_1}(\vec{R}_i) 
F^{*}(k_3,k_2;\omega ),
\nonumber\\
F^{*}(k_1,k_3;\omega )=F^{0*}(k_1;\omega )\delta (k_1-k_2)
+F^{0*}(k_1;\omega )\sum _{k_2,\vec{R}_i}U_{k_1k_3}(\vec{R}_i)
G(k_3,k_2;\omega )
\nonumber\\
+G^{0}(-k_1;-\omega )\sum_{k_2,\vec{R}_i}U_{-k_3-k_1}(\vec{R}_i)
F^{*}(k_3,k_2;\omega )
\nonumber\\
\label{e1}
\end{eqnarray}
with a similar equation for the function $F(k_1,k_3;\omega )$.
$U_{k_1k_2}(\vec{R}_i)$ 
is the matrix element of the impurity potential $U({\r}-\vec{R}_i)$ between two eigenstates (\ref{phi}). 
In high magnetic field we can assume that the scattering potential is weak compared to 
the separation between LL's (given by $\hbar \omega _c$).
Under these circumstances electrons scatter into the states belonging to 
the same LL. Therefore, in solving Eqs. (12) we will neglect 
the inter-Landau level scattering that becomes important only at much lower
fields.

We first solve Eqs. (\ref{e1}) in the Born approximation by summing
the diagrams that describe two consecutive electronic scatterings from the
same impurity. 
Since the impurity atoms are randomly distributed through the system, we 
have to average expressions (\ref{e1}) over the position of each impurity.
For a dilute concentration of scatterers with uncorrelated positions, 
we encounter two types of averages:
\begin{eqnarray}
<U_{k_1k_2}(\vec{R}_i)U_{k_2k_3}(\vec{R}_i)>_{\vec{R}_i}=
n_iU_0^2\frac{b_y}{L_xL_yL_z\sqrt{\pi }l}\delta _{k_{z1},k_{z3}}\delta _{\vec{q}_1,\vec{q}_3}\delta _{n_1,n_3}S_{n_1n_2}^N(\vec{q}_1-\vec{q}_2)
\nonumber\\
\label{snor}
\\
<U_{k_1k_2}(\vec{R}_i)U_{-k_3-k_2}(\vec{R}_i)>_{\vec{R}_i}=
n_iU_0^2\frac{b_y}{L_xL_yL_z\sqrt{\pi }l}\delta _{k_{z1},k_{z3}}
\delta _{\vec{q}_1,\vec{q}_3}\delta _{n_1,n_3}S_{n_1n_2}^A(\vec{q}_1,\vec{q}_2) 
\nonumber\\
\label{sanom}
\end{eqnarray}
where $<...>_{\vec{R}_i}$ denotes the average over the impurity positions and
$n_i$ is the impurity concentration. We find by inspection that $S_{n_1n_2}^N(\vec{q}_1-\vec{q}_2)$ in Eq. (\ref{snor}) is a weak function
of quasimomenta $\vec{q}_1-\vec{q}_2$ and LL indices $n_1$ and $n_2$. 
For $n_1+n_2\geq 5$, 
to a very good approximation, 
$S_{n_1n_2}^N(\vec{q})$ 
is independent of either 
$\vec{q}$ or $n$. 
The function $S^A_{n_1n_2}(\vec{q}_1,\vec{q}_2)$ in Eq. (\ref{sanom}) is obtained as:
\begin{equation}
S^A_{n_1n_2}(\vec{q}_1,\vec{q}_2)=\sqrt{2}
\sum_{k=0}^{min[n_1,n_2]}\frac{(2k-1)!!}{(2k)!!}f_{(n1-k)(n1-k)}(\vec{q}_1)f_{(n_2-k)(n_2-k)}^{*}(\vec{q}_2)
\label{ff}
\end{equation}
where matrix elements $f_{nn}({\q})=\Delta _{nn}({\q})/\Delta _0$ are 
calculated in (\ref{delta}) for the order parameter $\Delta ({\r})$ from  
the lowest LL of BCS pairs (\ref{order}).     

Using the above averages and the expressions (\ref{f}) we can bring the set of Eqs. (\ref{e1}) to the form:
\begin{eqnarray} 
\left( i\omega -\epsilon _n(k_z)-\Sigma ^N(\omega )\right) G_n(k_z,{\q};\omega )
+\left( \Delta _{nn}({\q})+\Sigma _{nn}^A({\q};\omega )\right) F_n^{*}(k_z,{\q};\omega )=1~, 
\nonumber\\
\left( i\omega +\epsilon _n(k_z)+\Sigma ^N(-\omega )\right) F_n^{*}(k_z,{\q};\omega )
+\left( \Delta _{nn}^{*}({\q})+\Sigma _{nn}^{A*}({\q},\omega )\right)G_n(k_z,
{\q};\omega )=0~, 
\nonumber\\
\label{born}
\end{eqnarray} 
The diagonal (normal) self-energy $\Sigma ^N(\omega )$ and off-diagonal
(anomalous) self-energy $\Sigma _{nn}^A({\q};\omega )$ can be expressed as:
\begin{eqnarray}
\Sigma ^N(\omega )=n_iU_0^2\frac{b_y}{L_xL_yL_z\sqrt{\pi }l}\sum_{m,k_z,{\k}}G_m(k_z,{\k};\omega ) 
\nonumber\\ 
\Sigma _{nn}^A({\q};\omega )=n_iU_0^2\frac{b_y}{L_xL_yL_z\sqrt{\pi }l}\sum_{m,k_z,{\k}}S^A_{nm}({\q},{\k})
F_m(k_z,{\k};\omega )
\label{A}
\end{eqnarray}
where the functions $G_m(k_z,{\k};\omega )$ and $F_m(k_z,{\k};\omega )$ are
found as a solution of Eqs. (\ref{born}) and can be written in the form:
\begin{eqnarray}
G_m(k_z,{\k};\omega )=\frac{-i\tilde{\omega }-\epsilon _m(k_z)}{\tilde{\omega }^2+
\epsilon _m^2(k_z)+|\tilde{\Delta }_{mm}({\k})|^2},
\nonumber\\ 
F_m(k_z,{\k};\omega 
)=\frac{\tilde{\Delta }_{mm}({\k})}{\tilde{\omega }^2+\epsilon _m^2(k_z)+|\tilde{\Delta }_{mm}({\k})|^2};
\label{gnew}
\end{eqnarray}
with
\begin{equation}
i\tilde{\omega }\equiv i\omega -\Sigma ^N(\omega ),~~~~~~\tilde{\Delta }_{mm}({\k})\equiv 
\Delta _{mm}({\k})+\Sigma _{mm}^A({\k};\omega )~. 
\label{om}
\end{equation}
The sums over $k_z$ in Eqs. (\ref{A}) can be readily done so that  
Eqs. (\ref{om}) become self-consistent equations for the self-energies:
\begin{equation} 
\tilde{\omega }=\omega +\Gamma _0\frac{1}{N(0)}\sum_{\nu =0}^{n_c}
\frac{m}{4\pi ^3k_{F\nu }}\int d{\k}\frac{\tilde{\omega }}{\sqrt{\tilde{\omega }^2+|\tilde{\Delta} _
{\nu \nu }({\k})|^2}} 
\label{om1}
\end{equation}
and
\begin{eqnarray}
\tilde{\Delta }_{nn}({\q})=\Delta _{nn}({\q})+
\frac{\Gamma _0}{N(0)}\sum_
{k=0}^n\frac{\sqrt{2}(2k-1)!!}{(2k)!!}f_{(n-k)(n-k)}({\q})
\nonumber\\
\times \sum_{\nu =k}^{n_c}
\frac{m}{4\pi ^3k_{F\nu }}\int d{\k}\frac{f^*_{(\nu -k)(\nu -k)}({\k})
\tilde{\Delta }_{\nu \nu }({\k})}{\sqrt{\tilde{\omega }^2+|\tilde{\Delta }_{\nu \nu }
({\k})|^2}}
\label{del1}
\end{eqnarray}
where $N(0)$ is the density of states at the Fermi level of the normal metal in
zero field. 
$\Gamma _0=\pi n_i
U_0^2N(0)$ is the scattering rate due to the disorder (defined in zero field).
It is assumed that $\Gamma _0/E_F\ll 1$ within the Born approximation. 

The exact self-consistent solution of Eqs. (\ref{om1}) and (\ref{del1}) 
is extremely difficult
to obtain due to the coupling of matrix elements $\Delta _{nn}({\q})$ with
$\Delta _{(n-k)(n-k)}({\q})$ in Eq. (\ref{del1}). Nevertheless,
this problem can be simplified if we notice that the behavior of $\Delta _{nn}
({\q})$ for different LL indices is very similar around the gapless points in 
Eqs. (\ref{epsilon}) and differs considerably only in the regions in the MBZ that are gapped by large $\Delta $. We are 
primarily interested in the low-energy behavior of the density of states
and low-temperature thermodynamic properties of a dirty superconductor in a
high magnetic field. These properties are governed by the quasiparticle
excitations around nodes in $\Delta _{nn}({\q})$. Therefore, retaining only the
$k=0$ term in Eq. (\ref{del1}) represents a reasonable approximation 
in solving Eqs. (\ref{om1}) and (\ref{del1}). Furthermore, we have demonstrated  
numerically  
that the sum of $k\neq 0$ terms in (\ref{del1}) is $\leq 10\%$ of
the $k=0$ term for small frequencies, while it is negligible for higher frequencies
(of order $\Delta $). Within this approximation $\tilde{\Delta }_{nn}({\q})=\tilde{\Delta }f_{nn}({\q})$ so that Eqs. (\ref{om1})
and (\ref{del1}) are combined as
\begin{equation}
u=\frac{\omega }{\Delta } +\zeta \frac{1}{N(0)}\sum _{n=0}^{n_c}\frac{m}
{4\pi ^3k_{Fn}}\int d{\k}\frac{u(1-\sqrt{2}|f_{nn}({\k})|^2)}
{\sqrt{u^2+|f_{nn}({\k})|^2}}
\label{comb}
\end{equation}
where $u=\tilde{\omega }/\tilde{\Delta }$ and $\zeta =\Gamma _0/\Delta $. The amplitude 
$\Delta \equiv \Delta (H,T,\Gamma )$ has to be determined from the self-consistent
equation: 
\begin{equation}
\Delta ({\r})=VT\sum _{\omega }F({\r},{\r};\omega )
\label{self3}
\end {equation}
If we take the order parameter $\Delta  
({\r})$ entirely in the lowest Landau level for the Cooper pair, 
Eq. (\ref{self3}) can be rewritten 
as
\begin{equation}
\Delta =\tilde{V}T\frac{\sqrt{2}b_y}{\sqrt{\pi }lL_xl_yL_z}\sum _{\omega }
\sum _{k_z{\q}n}\frac{f_{nn}^*({\q})\tilde{\Delta }_{nn}({\q})}{
\tilde{\omega }^2+\varepsilon (k_z)^2+|\tilde{\Delta }_{nn}({\q})|^2}
\label{self4}
\end{equation}
where $\tilde{V}(H,T)$ is the BCS pairing interaction that is, in general, renormalized by 
off-diagonal terms due to the coupling of electrons from LL's 
separated by $\hbar \omega _c$ or more \cite{thesis}. One can think of $\tilde{V}(H,T)$ as being chosen to reproduce the true self-consistent $\Delta
(H,T)$ in a formalism that keeps only the diagonal terms. Its explicit form is
easily computed in the $\Delta /\hbar \omega _c\ll 1$ regime \cite{thesis}. Note
that, once disorder is included, $\tilde{V}(H,T)$ {\em itself} has to be recomputed 
self-consistently. Here we ignore this complication on the grounds that a
modest sacrifice in quantitative accuracy (for $\Delta /\hbar \omega _c<1$)
is justified in the face of overwhelming numerical difficulty in determining
self-consistent $\Delta (H,T)$ in presence of disorder. 

The set of Eqs. (\ref{comb}) and (\ref{self4}) completely describes the effect of disorder
on the superconducting state within the Born approximation and enables us to
calculate various physical quantities.
The superconducting density of states in the presence of impurities is 
defined as
\begin{equation}
{\cal N}_s(\omega )=-\frac{1}{\pi }\Im m\int d{\r}{\cal G}({\r},{\r};i\omega)=
-\frac{1}{\pi L_xL_yL_z}\Im m
\sum _{n,k_z,{\q}}G_n(k_z,{\q};i\omega )|_{i\omega =
\omega +i\delta }
\label{dens2}
\end{equation}
where $G(k_z,{\q};i\omega )|_{i\omega =\omega +i\delta }$ is given by 
expression (\ref{gnew}) in which the analytic continuation to real
frequencies is performed. Equation (\ref{comb}) is an implicit
equation from which $u=u[\omega /\Delta ]$ is to be calculated. Once $u 
$ is known, the density of states (\ref{dens2}) can be obtained as
\begin{equation}
{\cal N}_s(\omega )/N(0)=\frac{1}{N(0)}\Im m\sum _{n=0}^{n_c}\frac{m}{4\pi ^3k_{Fn}}
\int d{\q}\frac{u}{\sqrt{|f_{nn}({\q})|^2-u^2}}
\label{dens3}
\end{equation}

In Fig. 1 we plot ${\cal N}_s(\omega )/N(0)$ for several values of the parameter $\zeta =
\Gamma _0/\Delta $ when $n_c=10$.  
Two kinds of behavior are present: For $\zeta \leq 0.9$
the superconducting density of states vanishes at the Fermi level and ${\cal N}_s(\omega )\sim 
\omega ^2$ for small $\omega $. This is the same behavior as one finds in 
a pure system, just that the coefficient in front of $\omega ^2$ is increased
from its clean system value.  When $\zeta >0.9$, a finite density of states is created at
the Fermi level, although of course it is smaller than in the normal state.
In this regime ${\cal N}_s(\omega )\sim {\cal N}_s(0)+{\rm const.}\times \omega $ for small $\omega $.
At higher energies, we observe that the peak in the density of states located at $\omega /
\Delta \approx 1/\sqrt{2}$ in the clean system is reduced and broadened by
disorder. As the impurity concentration (measured by the parameter $\zeta $) increases, the peak eventually disapeares.
Note, though, that our calculation might be less accurate at the higher 
energies (of the order $\Delta $) due to the number of approximations that are, as explained above, strictly applicable only at the low energies. 
 Furthermore, the true behavior of the peak 
can be investigated only if $\Delta =\Delta (\Gamma ,T)$ is found from
the self-consistent equation (\ref{self4}).
\section{T-matrix approximation}
The self-energies $\Sigma _{nn}^N({\q};\omega )$ and $\Sigma _{nn}^A({\q};\omega )$
of the superconducting system obeying Eqs. (\ref{born}) are closely
related to the diagonal (with respect to the Magnetic Translation
Group basis) T-matrix elements in a single-site approximation as
\begin{eqnarray}
\Sigma _{nn}^N({\k};\omega )=n_i<T^{11}(k,k;\omega )>_{\vec{R}_{i}} 
\nonumber\\
\Sigma _{nn}^A({\k};\omega )=-n_i<T^{12}(k,k;\omega )>_{\vec{R}_{i}}
\label{definition}
\end{eqnarray}
where $T^{ij}(k_1,k_2;\omega )$ are the coefficients in the T-matrix expansion
over the complete set of eigenstates $k\equiv (k_z,{\k},n)$
\begin{eqnarray}
{\cal T}^{11}({\r},{\r}~';\omega )=\sum _{k_1,k_2}\phi _{k_1}({\r})\phi _
{k_2}^*({\r}~')T^{11}(k_1,k_2;\omega )
\nonumber\\
{\cal T}^{12}({\r},{\r}~';\omega )=\sum _{k_1,k_2}\phi _{-k_1}({\r})\phi _
{k_2}({\r}~')T^{12}(k_1,k_2;\omega ). 
\label{ts}
\end{eqnarray}
The $2\times 2$ T-matrix $\hat{{\cal T}}({\r},{\r}~';\omega )$ obeys the 
Lippmann-Schwinger equations
\begin{equation}
\hat{{\cal T}}({\r},{\r}~';\omega )=U({\r})\delta({\r}-{\r}~')\hat{\sigma }_z+\int d{\r}_1U({\r})
\hat{\sigma }_z\hat{{\cal G}}({\r},{\r}_1;\omega )\hat{\cal{T}}({\r}_1,{\r}~';
\omega )
\label{lippman}
\end{equation} 
where $\hat{\cal G}$ matrix elements are given by (\ref{expansion}) and $U({\r})$ is 
the impurity potential (\ref{imp}). As in SCBA, we neglect
the inter-Landau level scattering on the basis that in high magnetic field the scattering potential
is much weaker than $\hbar \omega _c$. 
After averaging over the impurity position, Eqs. (\ref{lippman}) for
the dirty but homogeneous superconductor reduce to
\begin{eqnarray}
T^{11}_{nn}(k_z,{\k};\omega )=U_0+U_0\frac{b_y}{L_xL_yL_z\sqrt{\pi }l}\sum _{q_z,{\q},m}G_m(q_z,{\q};\omega )
T^{11}_{mn}(q_z,{\q};k_z,{\k};\omega )
\nonumber\\
+U_0\frac{b_y}{l_xL_yL_z\sqrt{\pi }l}\sum _{q_z,{\q},m}
S_{nm}^A({\k},{\q})F_m(q_z,{\q};\omega )T^{21}_{mn}(q_z,{\q};k_z,
{\k};\omega )
\label{ts1}
\end{eqnarray}
and
\begin{eqnarray}
T^{21}_{nn}(k_z,{\k};\omega )=-U_0\frac{b_y}{L_xL_yL_z\sqrt{\pi }l}\sum _{q_z,{\q},m}S_{nm}^{A*}({\k},
{\q})F_m^*(q_z,{\q};\omega )T^{11}_{mn}(q_z,{\q};k_z,{\k};\omega )
\nonumber\\
+U_0\frac{b_y}{L_xL_yL_z\sqrt{\pi }l}\sum _{q_z,{\q},m}G_m(-q_z,{-\q};-\omega )
T^{21}_{mn}(q_z,{\q};k_z,{\k};\omega ))
\label{ts2}
\end{eqnarray}
where $G_m(q_z,{\q};\omega )$ and $F_m(q_z,{\q};\omega )$ are written in the 
form (\ref{gnew}) and $S_{nm}^A({\k},{\q})$ is given by 
formula (\ref{ff}). 
Around the gapless points of the excitation spectrum (\ref{epsilon})
the second term in equation (\ref{ts1}) is very small compared to the
first term.  This can be deduced if one inspects expressions (\ref{ff})	and
(\ref{ts2}) around the nodes of the excitation spectrum (\ref{epsilon}) in the MBZ. Keeping this conclusion in
mind, equations (\ref{ts1}) and (\ref{ts2}) can be solved
as
\begin{equation}
T^{11}(\omega )=\frac{\frac{b_y}{L_xL_yL_z\sqrt{\pi }l}\sum _{q_z,{\q},m}
G_m(q_z,{\q},m)}{1/U_0^2-\left[\frac{b_y}{L_xL_yL_z\sqrt{\pi }l}
\sum _{q_z,{\q},m}G_m(q_z,{\q},m)\right]^2}
\label{ts3}
\end{equation}
and
\begin{equation}
T^{21}_{nn}({\k};\omega )=-\frac{\frac{\sqrt{2}b_y}{L_xL_yL_z\sqrt{\pi }l}
f_{nn}^*({\k})\sum _{q_z,{\q},m}f_{mm}({\q})F_m^*(q_z,{\q};\omega )}
{1/U_0^2-\left[\frac{b_y}{L_xL_yL_z\sqrt{\pi }l}\sum _{q_z,{\q},m}
G_m(q_z,{\q},m)\right]^2}
\label{ts4}
\end{equation}
where we used the explicit form (\ref{ff}) for  
$S_{nm}^A({\k},{\q})$ (taking only the $k=0$ term in Eq. (\ref{ff}), see the discussion
in previous section). With the help of definitions (\ref{om}) and (\ref{definition}), the set of equations (\ref{ts3}) and (\ref{ts4}) can
be brought into the form
\begin{equation}
u=\frac{\omega }{\Delta }+\zeta \frac{\sum _n\frac{m}{4\pi ^3k_{Fn}N(0)}\int 
d{\q}(1-\sqrt{2}|f_{nn}({\q})|^2)u/\sqrt{u^2+|f_{nn}({\q})|^2}}
{c^2+\left[\sum_n\frac{m}{4\pi ^3k_{Fn}N(0)}\int d{\q}u/\sqrt{
u^2+|f_{nn}({\q})|^2}\right]^2} 
\label{unit}
\end{equation}
where $\zeta =\Gamma /\Delta $ and $u=\tilde{\omega }/\tilde{\Delta }$. 

Disorder is 
characterized with two parameters: $\Gamma =n_i/N(0)\pi =(n_i/n)E_F$, 
which measures 
the concentration of impurities $n_i$ relative to the electron density
$n$, and $c=1/\pi N(0)U_0$, which measures 
the strength of the scattering potential. The normal state inverse 
scattering rate $1/2\tau $ is found by taking $f_{nn}({\q})=0$
in (\ref{unit}) and letting $\omega \rightarrow 0$. This procedure 
yields $1/2\tau =\Gamma /(1+c^2)$, the result first obtained in Ref. 17 in the study of the transport properties of normal metals in a high
magnetic field.
The weak scattering limit is approached when $c^2$ is much larger than the
second term in the denominator of expression (\ref{unit}), while the strong scattering
limit is achieved when $c^2=0$. In the strong scattering limit, the approximation in which 
the inter-Landau level scattering is neglected eventually becomes unphysical, unless the magnetic field is so high that 
only the lowest LL is occupied. 

As in previous section, the superconducting density of states in the presence
of impurities is found from Eq.
(\ref{dens3}) once 
the solution $u=u[\omega /\Delta ]$ of Eq. (\ref{unit}) is found. Figures 2, 3 and 4 show how the superconducting density of states 
${\cal N}_s(\omega )/N(0)$ behaves as a function of the energy parameter 
$\omega /\Delta $ as one crosses from a weak scattering limit, $c=1.0$, (Fig.
2) to a strong scattering limit, $c=0.0$, (Fig. 4). For each value of $c$
(measuring the scattering strength) we present how the density of states changes as the impurity concentration, 
i.e., parameter $\zeta $ increases. There are two types of behavior present 
in figures. When $\zeta < c^2$ the density of states vanishes at the
Fermi level with ${\cal N}_s(\omega )\sim \omega ^2$ for small $\omega $,
the coefficient in front of $\omega ^2$ being increased from the clean
system value. When $\zeta \approx c^2$, the density of states still 
vanishes at the Fermi level, but ${\cal N}_s(\omega )\sim \omega $ for
small $\omega $. Further increase of the concentration $n_i$ such
that $\zeta >c^2$ creates a finite density of states ${\cal N}_s(0)$ at the Fermi
level. In the strong scattering limit, $c\ll 1$, the superconducting density of states is finite at
the Fermi level for any non-zero concentration of impurities. 
In this limit, ${\cal N}_s(0)/N(0)\approx 2(\gamma /\Delta )$
where $\gamma =\Delta \sqrt{\zeta /2}$ for $\zeta \ll 1$.
Furthermore, below $\omega /\Delta =\sqrt{\zeta}$ in the strong scattering limit,
a peak is observed centered at zero energy. This peak suggests  
formation of a quasi-bound, resonant state which is analog of
a Shiba state formed in the energy gap of a conventional s-wave
superconductor, as a result of multiple scattering off a magnetic
impurity \cite{Shiba}. As one moves away from the strong scattering limit, 
this zero-energy peak disappears.
At higher energies, we find a similar behavior of the density of 
states to the one observed in the Born approximation of the previous section:
The peak, located at $\omega /\Delta \approx 1/\sqrt{2}$ in the clean
system, is reduced 
and broadened as the impurity concentration (measured by parameter $\zeta $) 
increases. Also, this peak is slightly shifted to higher $\omega /\Delta $
as $\zeta $ increases, suggesting a stronger reduction in BCS amplitude
$\Delta =\Delta (T,\Gamma )$ than what is found in the Born approximation.
\section{Conclusions}
In this paper we have analyzed the influence of a dilute static disorder on 
superconducting properties in a high magnetic field. We considered a 
dirty but homogeneous superconductor for which the order parameter $\Delta ({\r})$ is not
influenced by the presence of the impurities and still forms the perfect
Abrikosov triangular lattice. We considered the weak-scattering limit
within a self-consistent Born approximation while the strong-scattering limit
was treated within a T-matrix approximation for superconducting 
self-energies.

We found that for small impurity concentrations and weak scattering 
potentials the superconducting density of states behaves as $E^2$ for small
energies $E$, the same behavior as that found for the pure superconductor in a magnetic
field \cite{sasa}. When disorder becomes stronger than some critical value, a finite
density of states (but still smaller then the normal state value) is
created at the Fermi level. The finite superconducting density of states 
at the Fermi level signals the broadening of gapless points into 
gapless regions in the MBZ. 
It is interesting to mention that 
this behavior is similar to that of dirty superfluid $^3$He 
films \cite{vallis}, and somewhat similar to the
behavior of the density of states in anisotropic heavy-fermion superconductors
\cite{hirch}.

The experimental property of a superconductor in which  the absence
of a quasiparticle gap over some region of the Fermi surface will be most obviously
felt is the specific heat. In a clean system in high magnetic
field at low temperatures $c_v\sim AT^3$, where $A$ is the 
field dependent coefficient \cite{sasa}.
In a dirty but homogeneous superconductor, instead
of the $T^3$ law one finds linear behavior at low temperatures
with the coefficient reduced by the factor $\sim 2(\gamma /\Delta )$ from
the normal state value. Detailed measurements of heat capacity at 
very low temperatures and high magnetic fields are not yet found  
in the literature. We propose a class of A-15 superconductors as 
good candidates in which the linear temperature law of heat capacity
at high magnetic field can be discovered. These systems have 
experimentally accessible upper critical fields and are clear examples of materials for which the LL
quantization in high fields plays an important role \cite{prl,corcoran,spring}.

This work has been supported in part by the NSF Grant No. DMR-9415549.

\newpage
\begin{figure}
\caption{Quasiparticle density of states ${\cal N}_s(\omega )/N(0)$ vs 
reduced energy $\omega /\Delta $ in the Born approximation as a function
of parameter $\zeta =\Gamma _0/\Delta $.}
\label{plot0}
\end{figure}
\begin{figure}
\caption{Quasiparticle density of states ${\cal N}_s(\omega )/N(0)$ vs reduced 
energy $\omega /\Delta $ when $c=1.0$,
as a function of the impurity concentration parameter
$\zeta =\Gamma /\Delta $.} 
\label{plott1}
\end{figure}
\begin{figure}
\caption{Quasiparticle density of states ${\cal N}_s(\omega )/N(0)$ vs reduced 
energy $\omega /\Delta $ when $c=0.5$,
as a function of the impurity concentration parameter
$\zeta =\Gamma /\Delta $.} 
\label{plott2}
\end{figure}
\begin{figure}
\caption{Quasiparticle density of states ${\cal N}_s(\omega )/N(0)$ vs reduced 
energy $\omega /\Delta $ when $c=0.0$,
as a function of the impurity concentration parameter
$\zeta =\Gamma /\Delta $.} 
\label{plott3}
\end{figure}
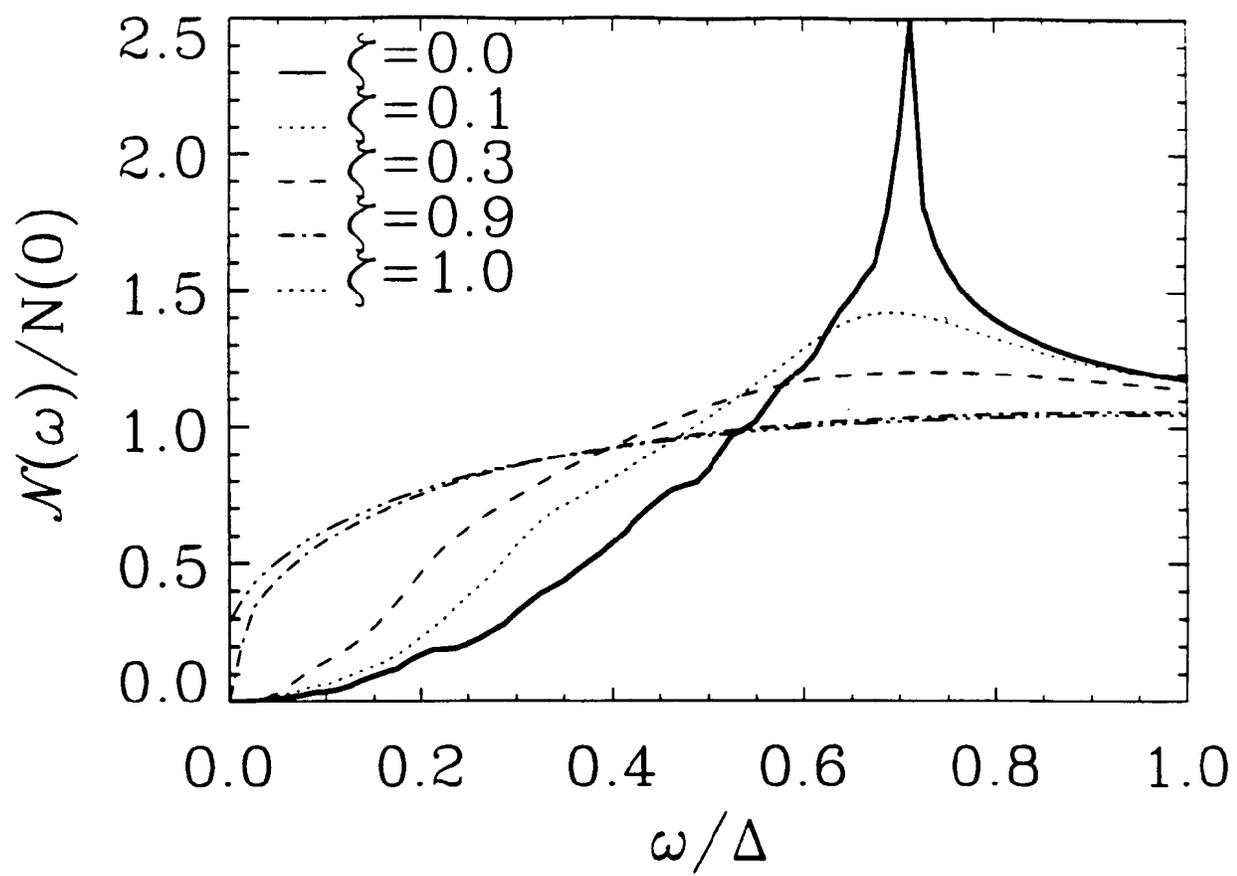
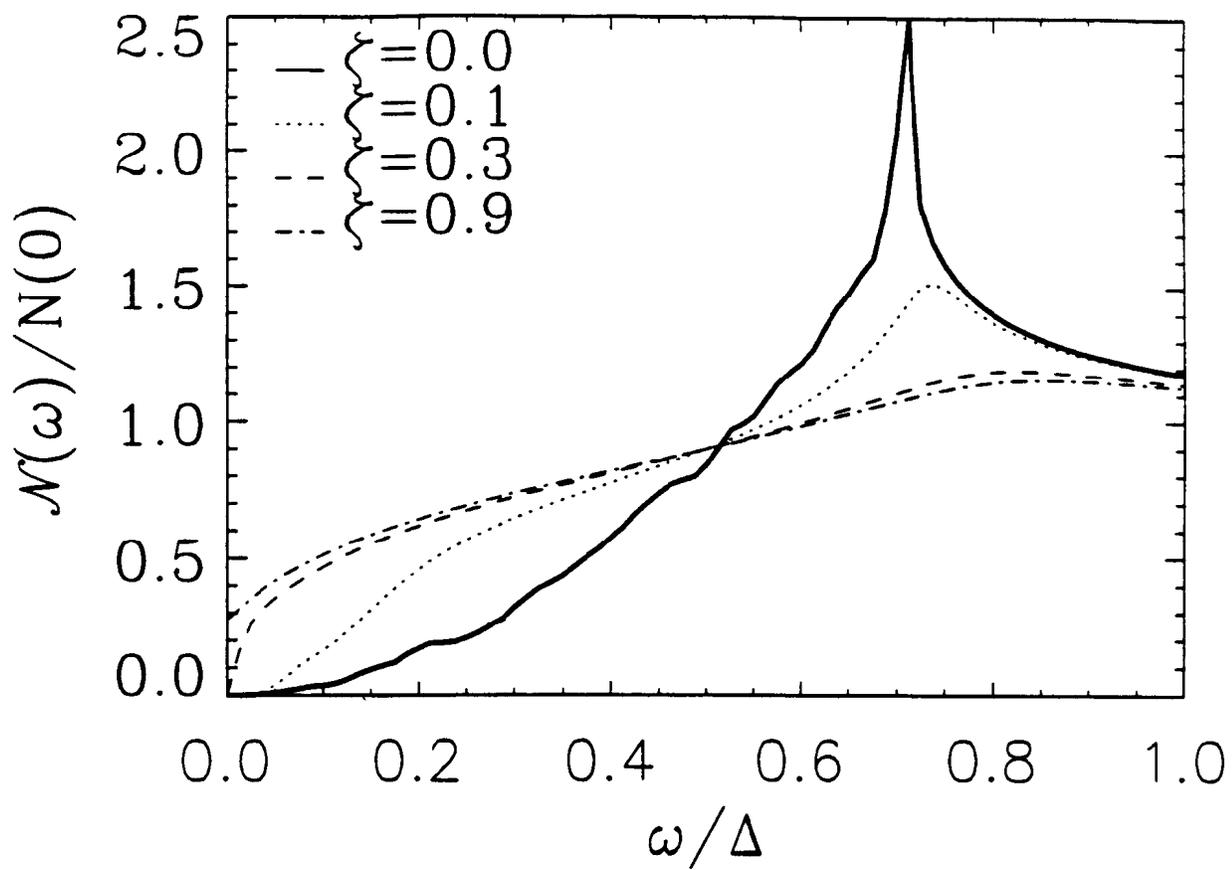
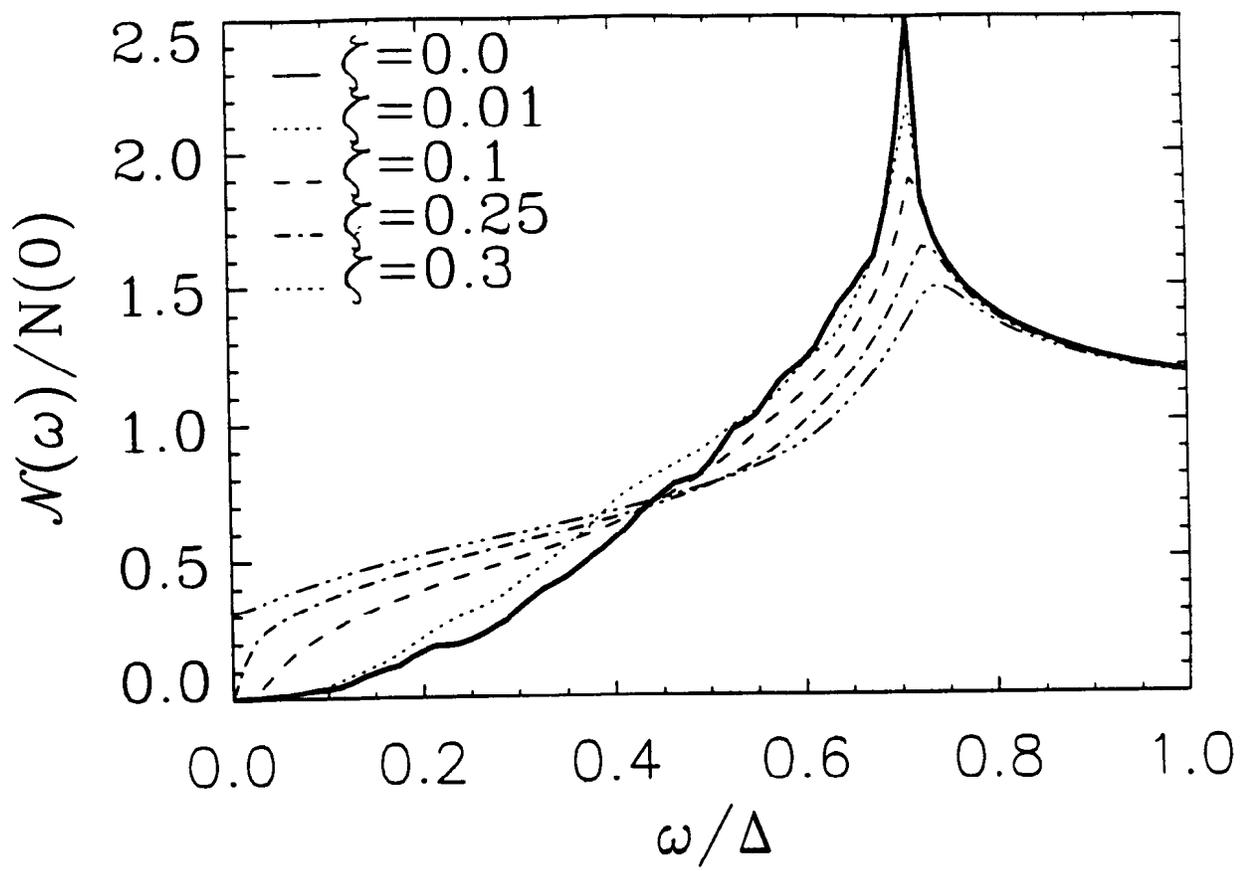
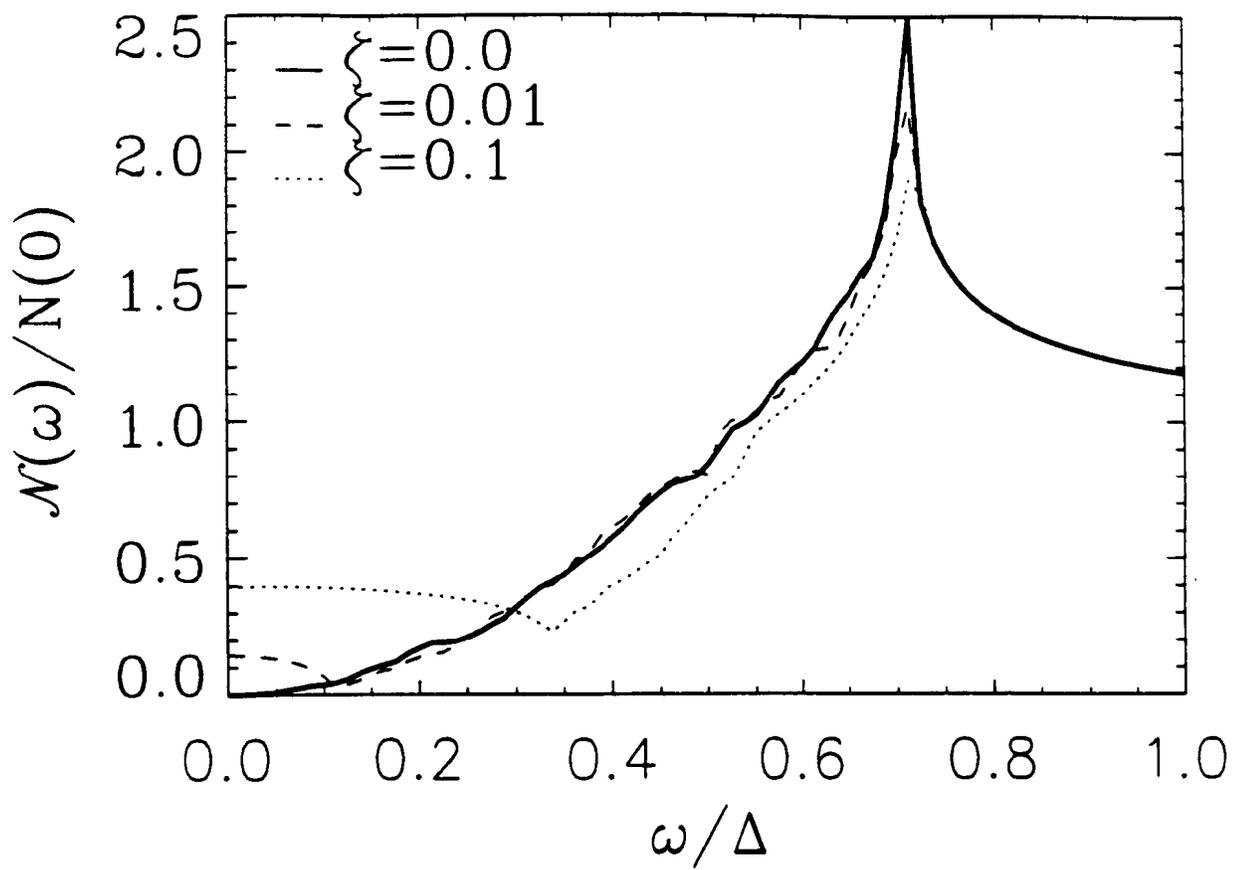
\end{document}